\documentclass[conference]{IEEEtran}
\IEEEoverridecommandlockouts
\usepackage{cite}
\usepackage{amsmath,amssymb,amsfonts}
\usepackage{algorithmic}
\usepackage{float}
\usepackage{graphicx}
\usepackage{textcomp}
\usepackage{xcolor}
\usepackage{makecell}
\def\BibTeX{{\rm B\kern-.05em{\sc i\kern-.025em b}\kern-.08em
    T\kern-.1667em\lower.7ex\hbox{E}\kern-.125emX}}
\begin{document}

\title{GenAI Security: Outsmarting the Bots with a Proactive Testing Framework\\
}

\author{\IEEEauthorblockN{Sunil Kumar Jang Bahadur}
\IEEEauthorblockA{\textit{AI \& GenAI Specialist} \\
\textit{Cloud GTM}\\
\textit{Google}\\
Mumbai, India \\
sjangabahadur@google.com}
\and
\IEEEauthorblockN{Gopala Dhar}
\IEEEauthorblockA{\textit{AI Engineer, AI Services} \\
\textit{Google Cloud Consulting (GCC)}\\
\textit{Google}\\
Mumbai, India \\
gopalad@google.com}
\and
\IEEEauthorblockN{Lavi Nigam}
\IEEEauthorblockA{\textit{Developer Relations Engineer} \\
\textit{Cloud AI and Industry Solutions} \\
\textit{Google}\\
Gurugram, India\\
lavinigam@google.com}
}

\maketitle

\begin{abstract}
The increasing sophistication and integration of Generative AI (GenAI) models into diverse applications introduce new security challenges that traditional methods struggle to address. This research explores the critical need for proactive security measures to mitigate the risks associated with malicious exploitation of GenAI systems. We present a framework encompassing key approaches, tools, and strategies designed to outmaneuver even advanced adversarial attacks, emphasizing the importance of securing GenAI innovation against potential liabilities. We also empirically prove the effectiveness of the said framework by testing it against the SPML Chatbot Prompt Injection Dataset. This work highlights the shift from reactive to proactive security practices essential for the safe and responsible deployment of GenAI technologies.
\end{abstract}


\section{Introduction}
The rapid advancement and widespread adoption of Generative Artificial Intelligence (GenAI) models have opened up exciting new possibilities across various domains. However, this progress has also brought few significant security concerns, notably the vulnerability of these models to adversarial attacks\cite{b1}. Among these, prompt injection attacks pose a critical threat, allowing malicious actors to manipulate the behavior of GenAI models for unintended and often harmful purposes. Traditional security approaches, reliant on reactive measures, struggle to keep pace with the evolving landscape of prompt injection techniques.
A proactive testing framework that identifies and mitigates vulnerabilities, leverages the principles of Red Teaming and Blue Teaming within a unified and automated pipeline, is crucial. This framework will not only strengthen the security posture of GenAI systems but also provides valuable insights into the evolving sophistication of prompt injection attacks, empowering developers and security professionals to build more resilient and trustworthy AI solutions for the safe and responsible deployment of GenAI technologies.

\section{Approach}
Red teaming and Blue teaming are used in cyber-security to identify vulnerabilities and strengthen their defenses through a controlled adversarial process. "Red teaming" refers to a simulated cyber-attack performed by a group of security experts to identify vulnerabilities in a system, while "Blue teaming" refers to the defensive team responsible for detecting and responding to these simulated attacks. 
\begin{figure}[H]
    \centering
    \includegraphics[width=\linewidth]{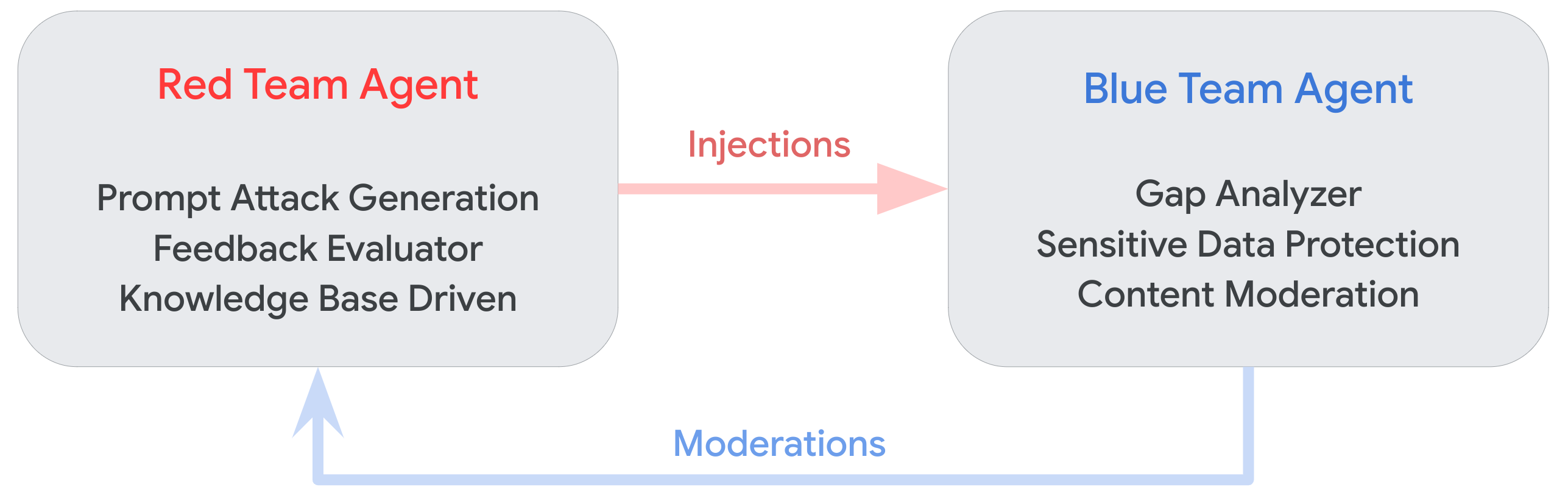} 
    \caption{Proactive testing framework approach}
    \label{fig1}
\end{figure}
However, given the rapid release of new GenAI models and the constant discovery of new bypass techniques, traditional Red Teaming and Blue Teaming approaches face several limitations, some of those are listed below:
\begin{itemize}
\item \textbf{Stagnates Fast}: Traditional approaches provide solutions at a snapshot in time; such a security posture can quickly become outdated.
\item \textbf{Difficult to Scale}: These solutions are challenging to apply across large and complex organizations.
\item \textbf{Limited Scope}: These methods often are focused on specific systems or attack vectors, leaving other areas unexplored.
\item \textbf{Resource Intensive}: Traditional methodologies requires significant time, budget, and specialized personnel.
\item \textbf{Reactive}: These primarily identify existing vulnerabilities, and not emerging threats.
\item \textbf{Human Bias}: Traditional methodologies are prone to human error and subjective assessments.
\end{itemize}
The proposed solution involves developing a proactive testing framework that utilizes GenAI models to drive both "Red Teaming"\cite{b2} and "Blue Teaming" agents. The "Red Teaming" agent generates prompt attacks based on existing knowledge and continuous feedback, while the "Blue Teaming" agent analyzes successful prompt injections to identify vulnerabilities and recommend corrective actions.

\section{Architecture}
The core concept adopted for constructing a proactive testing framework for GenAI applications is an agentic approach leveraging the GenAI models as the brain. The architecture will consist of two primary agents, viz.:\\

\textbf{"Red Teaming" agent}: The "Red Teaming" agent's objective is t\o investigate various methods to inject malicious instructions into user prompts while circumventing safety filters and avoiding suspicion. It utilizes a Knowledge Base built upon known and newly discovered prompt attack techniques and research like SurrogatePrompt\cite{b3}, SneakyPrompt\cite{b4}. Additionally, it employs a shared memory to retain learnings for both short-term and long-term use in specific scenarios.\\

\textbf{"Blue Teaming" agent}: The “Blue Teaming” agent's goal is to examine the results from the Red Teaming agents, identify weaknesses in the existing security or workflow process, and suggest a list of services that can mitigate the attacks, based on the available list of services.\\
\begin{figure}[H]
    \centering
    \includegraphics[width=\linewidth]{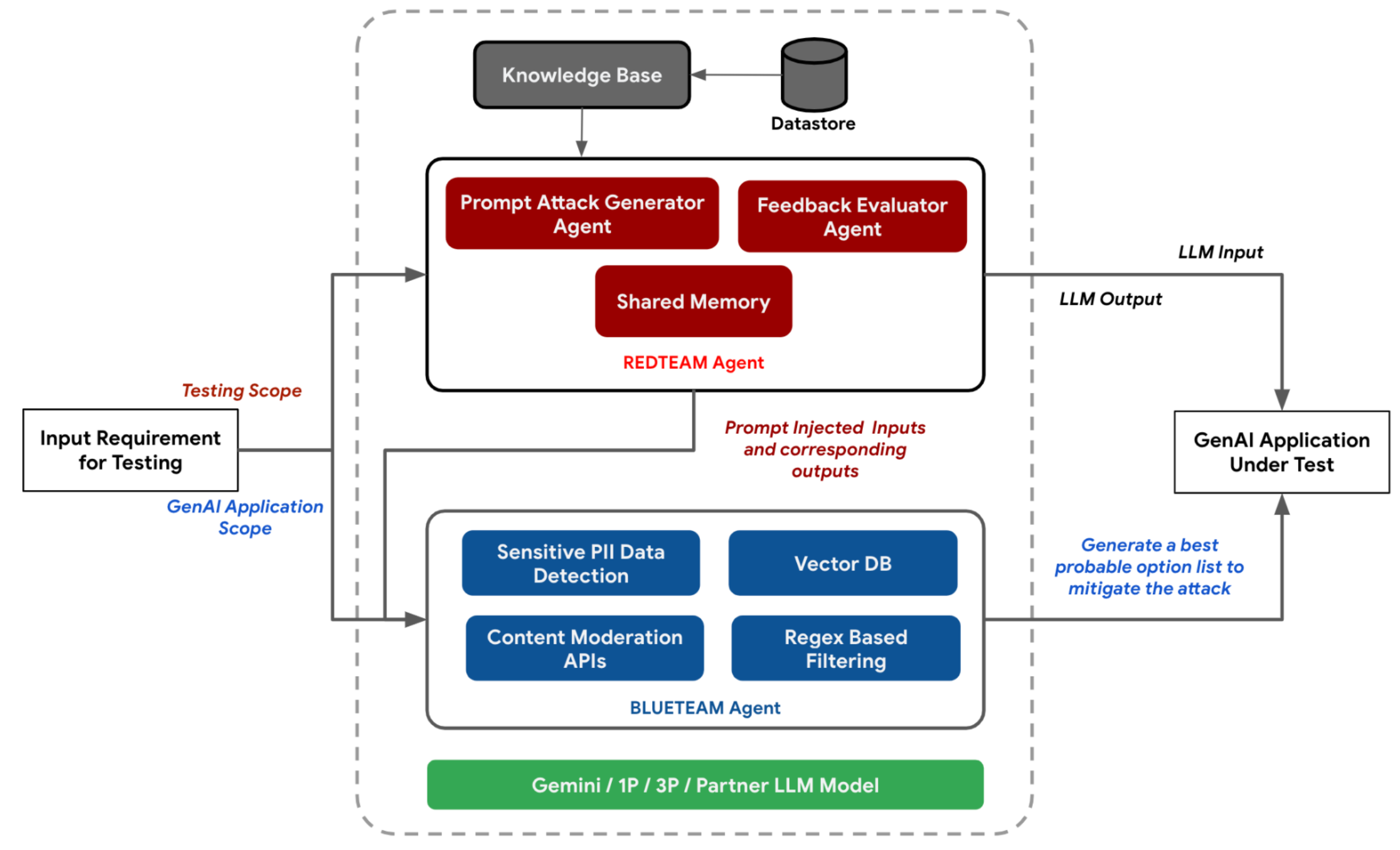} 
    \caption{GenAI testing framework architecture}
    \label{fig2}
\end{figure}
In future work, the collected data and scenarios can be later used to further fine-tune the all models for better understanding and performance.
Given a particular use case and specific vulnerabilities that need to be tested, the "Red Teaming" agent will generate and investigate potential scenarios as part of an end-to-end workflow. The agent will then share its findings with the "Blue Teaming" agent, who will recommend solutions to address any identified gaps.\par 
The feedback evaluator will continuously challenge the "Red Teaming" agent to produce increasingly optimized and complex prompt injection attacks if the existing workflow, utilizing the constantly updated Knowledge Base, can successfully defend against them. Additionally, the "Blue Teaming" agent can be expanded to incorporate more customized options to address the identified vulnerabilities.
\section{Experimentation}

\subsection{Data}
The SPML Chatbot Prompt Injection Dataset\cite{b5}, comprising 16K records, was employed to facilitate our experimentation. This dataset encompasses a substantial collection of system prompts crafted to simulate authentic chatbot interactions, complemented by a diverse assortment of annotated user prompts aimed at executing prompt injection attacks. It includes "System Prompts" and "User Prompts," accompanied by a binary prompt injection indicator utilized to evaluate the Red Teaming agent. Furthermore, the "Degree" field, denoting the severity of prompt injection on a numerical scale from 0 (no injection) to 10 (highest severity), is leveraged to assess our proactive testing framework.

\subsection{Approaches and Challenges Faced}

In order to evaluate the Red Teaming agent, with the given dataset, we needed to formulate a strategy that vets the Red Teaming agents prompts. Due to the limitations of the open source datasets available as well as evaluation methodology to benchmark generated injections, we use a proxy in order to evaluate the agent.\par
Since large language models can practically generate any number of variations of an "injected" prompt, we utilize an evaluator model that takes the input prompts of the Red Teaming agent and then validates whether or not the agent is able to categorize a "User Prompt" as an injection for a given "System Prompt". Essentially we transform the evaluation problem statement to a binary classification problem, in order to quantify the effectiveness of the Red Teaming agent. The evaluator model used in our case was Google's gemini-1.5 model.\par
In order to evaluate the blue teaming  agent, we utilize the "Degree" field from the dataset and compare it against the number of recommendations made by the Blue Teaming agent. Each recommendation is a suggestion made by the Blue Teaming agent to mitigate the attempted injection. The hypothesis being that, for a severe prompt injection attack with a higher "Degree" value the number of recommendations generated by the agent should be equally high. In our experimentation prompt for the Blue Teaming agent, we provided a list of 4 recommendations for the agent to choose from. Hence the maximum recommendations value can be 4.

\subsection{Evaluation \& Results}
In order to quantify the evaluation metrics for the Red Teaming agent, we utilize the metrics used to validate any binary classification model. 

\begin{table}[H]
\caption{Experimentation Results for Red Teaming Agent}
\begin{center}
\renewcommand{\arraystretch}{2} 
\begin{tabular}{|c|c|}
\hline
Metric & Value\\
\hline
Accuracy & 0.9767\\
\hline
Precision & 0.9951\\
\hline
Recall & 0.9751\\
\hline
F1 Score & 0.9850\\
\hline
\end{tabular}
\label{table1}
\end{center}
\end{table}
In Table I, we observe that the Red Teaming agent performs significantly well in terms of observing possible injections. With a F1 score of 0.985, the agent has produced reasonably limited false positives and false negatives. We also observe that the Red Teaming agent's evaluation has a higher precision than recall, implying that there are certainly more false negatives than there are false positives, in other words, there are more scenarios wherein the Red Teaming agent falsely identifies an injected scenario as not an injection. Upon further inspection of such examples, we realized that it happens in the cases where the "Input Prompt" is highly nuanced and the "User Prompt" ever so slightly nudges the model to deviate from its instructions. For eg: A case wherein the "System Prompt" directs the model to be a helpful medical assistant which does not recommend or prescribe medicines, combined with an injected "User Prompt" that forces the model to recommend a medicine for a life threatening condition, is considered "Injection" in the ground truth, however the Red Teaming agent fails to classify and understand this as an injected prompt.\par
For the Blue Teaming agent's evaluation, we compare the severity of injection present in the dataset with the amount of preventive recommendations suggested by the Blue Teaming agent. 
\begin{figure}[H]
    \centering
    \includegraphics[width=\linewidth]{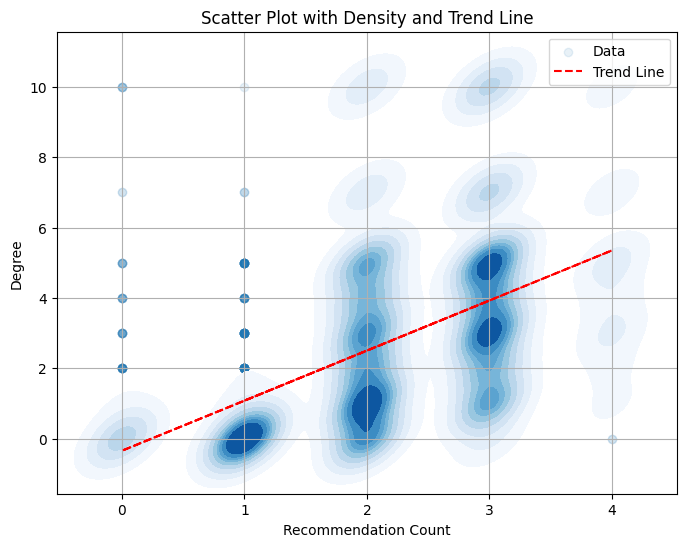} 
    \caption{Blue Teaming Agent Evaluation Plot}
    \label{fig3}
\end{figure}
\begin{table}[H]
\caption{Confusion Matrix for Red Teaming Agent Evaluation} 
\begin{center}
\renewcommand{\arraystretch}{3} 
\begin{tabular}{|c|c|c|}
\hline
& \makecell{Predicted \\ No Injection} & \makecell{Predicted \\ Injection} \\
\hline
\makecell{Ground Truth \\ No Injection} & 3410 & 60 \\
\hline
\makecell{Ground Truth \\ Injection} & 312 & 12221 \\
\hline
\end{tabular}
\label{table2} 
\end{center}
\end{table}
Upon plotting a density scatter plot of the recommendations suggested by the Blue Teaming agent vs the degree of injection severity, in Fig.~\ref{fig3}, we observe, the count of recommendations suggested by the agent to mitigate the prompt injection increases as the degree of prompt injection increases. The trend can be observed through the red dotted line in the same figure.

\section{Conclusion}
This research has presented a proactive security framework for GenAI applications, emphasizing a shift from reactive to preemptive measures in the face of evolving adversarial threats. The framework, which incorporates both red teaming and blue teaming strategies, has demonstrated its effectiveness in identifying and mitigating potential vulnerabilities. Through experimental evaluations, it was observed that the Red Teaming agent achieved a high F1 score of 0.985, demonstrating its proficiency in identifying injection attacks with minimal false positives or negatives. The Blue Teaming agent showed a clear understanding of the severity of prompt injections, exhibiting a strong positive correlation between the complexity of the attack and the number of mitigation recommendations provided. These findings strongly suggest that the proposed agentic framework approach to GenAI application testing is robust, balanced, and well-suited for real-world applications. By adopting this framework, developers and organizations can better secure their GenAI systems, ensuring their safe and responsible deployment, thereby minimizing potential liabilities.

\end{document}